\documentclass[aps,pra,preprint,showpacs,groupedaddress,amsmath,amssymb,amsfonts]{revtex4-1}
\usepackage{graphicx}
\usepackage{bm}

\begin{document}


\title{Phase Transitions, Renormalization \\and Yang-Lee Zeros in Stock  Markets}


\author{J. L. Subias}
\email[e-mail for correspondence to the author:\\]{jlsubias@unizar.es}
\affiliation{Departamento de Ingenieria de Diseno y Fabricacion, Universidad de Zaragoza,\\C/Maria de Luna 3, 50018-Zaragoza, Spain}


\date{April 18, 2015}

\begin{abstract}
The present paper analyses the formal parallelism existing between the laws of thermodynamics and some economic principles. Based on previous works, we shall show how the existence in Economics of principles analogous to those in thermodynamics involves the occurrence of economic events that remind of well-known phenomenological thermodynamic paradigms (i.e., the magnetocaloric effect and population inversion). We shall also show how the phase transition and renormalization theory provides a natural framework to understand and predict trend changes in stock markets. Finally, current negotiation strategies in financial markets are briefly reviewed.
\end{abstract}


\maketitle

\section{Introduction}
One of the most interesting findings in Science is duality -i.e., the fact that there are laws formally identical in two different fields of the same discipline and that their principles are simply equivalent, thus requiring mere substitution of each concept by its corresponding pair. For instance, there are dual spaces in Geometry, so that -if dual concepts are interchanged in a true sentence- the corresponding dual sentence shall be obtained, and shall also be true. For example, the sentence "three non-aligned points determine a plane" find its dual sentence in "three non-parallel planes determine a point". Plane and point are dual concepts. However, the most remarkable aspect is that duality can occur between widely differing fields in a discipline. Let us consider L. Onsager's widely-known example. Onsager perspicaciously realized that the Langevin equation could satisfactorily describe a completely different phenomenon from that it had initially been conceived for -he realized that changing names in the Langevin equation leads to an equation that satisfactorily solves an apparently unrelated problem in statistical physics.
At this point, the question we pose is: can duality occur between completely different disciplines, so that it even transcends the reach itself of Physics and Mathematics?
If there is duality between fundamental principles, all laws, theorems, equations, etc. would then admit other formulations that would only involve changing every concept in them by its corresponding pair. Can this duality occur between such different disciplines as Physics and Economics?
A fundamental law of Physics usually taught in high school states that \emph{"Energy can be neither created nor destroyed, but it can only change form"}. In Economics there is a dual principle to the previous one: \emph{"Money can be neither created nor destroyed, it can only change pocket"}.
On the other hand, the second law of thermodynamics -which states that the entropy of an isolated system never decreases- is known to have universal validity and to apply in Biology, Ecology, Sociology and, of course, Economics.
Perhaps the most difficult aspect to "grasp" is that, in Economics, the \emph{"no arbitrage"} principle also has its counterpart. Indeed, consider the following statement -\emph{"if between markets A and B there is no arbitrage possibilities, nor between markets B and C, therefore there cannot be arbitrage possibilities between markets A and C either"}. When analysed in depth, this is nothing but the zeroth law of thermodynamics applied to Economics.
If the three fundamental laws of thermodynamics also apply in Economics, it is then logical to think that numerous economic phenomena will have their corresponding thermodynamic equivalent. The reason why these dualisms have not been observed so far is that most part of the task is still to be done: namely, translating physical variables (temperature, pressure, etc.) into economic variables, as Onsager did with \emph{generalized forces, generalized coordinates}, etc.
The present work summarises the findings obtained in previous works on phenomenological thermodynamics paradigms that have counterparts in Economics (namely, population inversion and magnetocaloric effect). Then, the latest findings on phase transitions, renormalization and Yang-Lee zeros in the financial context shall be described.
\section{Effective temperatures}
As a starting point for the present study, we shall adopt the definitions -applied to the field of financial markets- of the following variables: temperature, energy, magnetic field and entropy, as formulated in \cite{Subi13}.
Let us imagine the following context: a mixture of economic agents in perpetual interaction. A Brownian particle is suspended in this mixture, and randomly impacted by the numerous economic agents, describing a geometrical, one-dimensional Brownian motion. The foregoing is a possible conception of a financial market -i.e., the \emph{time series} of stock prices is conceived as the movement of a hypothetical Brownian particle hit by economic agents. This particle is then equivalent to a thermometer that allows us to measure the \emph{effective temperature} of the system it is suspended in. Such effective temperature is an estimation of \emph{market temperature} and can be calculated by (5) in [1]. Some cases in which the temperature sequence imitated well-known processes in phenomenological thermodynamics are described in \cite{Subi13} and shall be summarised next.
\subsection{Population inversion}
\begin{figure}[tb]
 \includegraphics*[width=\textwidth]{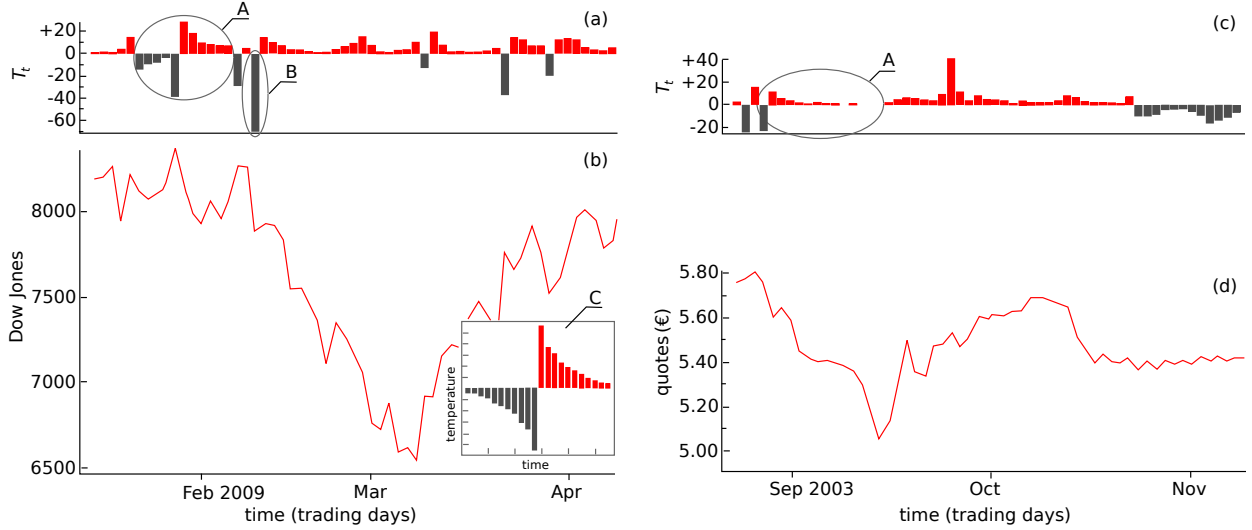}
  \caption{\label{Dow}~(a) Time series of temperatures $T_{t}$ corresponding to (b) the Dow Jones index. The inset~$C$ shows how a discrete register of a physical population inversion would be. Compare the inset with detail~$A$. ~(c) Time series of temperatures $T_{t}$ corresponding to (d) quotes of a particular stock. After a dividend payout, this stock suffers an extreme \textquotedblleft cooling'' near to $T_{t} = 0$ (detail $A$).}
 \end{figure}

Figure ~\ref{Dow}~(a) shows a type of graph that imitates the phenomenon known as population inversion in statistical physics. In this phenomenon, the external field is inverted so brusquely that spins cannot adapt to such a sudden commutation. This leaves the system in a highly unstable state. For some time, spins shall try to adapt to a new stable equilibrium. During this transition, temperature firstly rises and then falls, thus corresponding to a theoretical step from $T =-\infty$ to $T=+\infty$ \cite{Path11}. Note the dramatic fall of stock prices that takes place in the following days.
\subsubsection{Explanation of this phenomenon in economic terms}
The emergence of some breaking bad news that go against the status at the time being (the magnetic field is \emph{brusquely inverted}) leads investors (\emph{spins}) to instantly change their mind (very brief \emph{spin-field} interaction). However, selling their parcels of shares will take them much longer (much longer \emph{spin-network} interaction). The difference between these two typical times is what favours the population inversion phenomenon. Some hurry to sell their shares, and market temperature then turns negative and sharply falls towards $-\infty$, and then comes back from $+\infty$, decreasing and eventually becoming stable. Nevertheless, full balance has not been reached yet, since the sectors reacting first are those that own \emph{privileged information} -i.e., those that know the bad news before they break into the media. In the following days, the remaining parts of the markets will react and stock prices will slump.
\subsection{Magnetocaloric effect}
In a different environment, Fig. ~\ref{Dow}~(c) shows the case of a stock security whose investors are 'aligned' awaiting the payment of abundant dividends. This expectation represents a strong \emph{magnetic field} that suddenly vanishes the day after dividend payment. At this time, stock prices undergo 'cooling' close to $T=0$, imitating the magnetocaloric effect (detail A).
\subsubsection{Explanation of this phenomenon in economic terms}
Investors (\emph{spins}) expect to be paid abundant dividends: almost nobody sells or buys (spins are mainly aligned under a powerful \emph{magnetic field}). On the scheduled date, payment is made (the magnetic field suddenly \emph{disappears}) and investors randomly decide to keep or sell their shares after being paid their corresponding dividends. Thus, it shifts from a very \emph{ordered} situation to a very \emph{disordered} one with no money provision (\emph{energy}), since the remaining agents in the market know that dividends are to be \emph{deducted} from stock prices and, accordingly, stay out of it. Therefore, disordered state can only be reached through interaction at internal level: some buy what others sell and 'temperature' drops sharply (not market temperature, but that of these specific shareholders). The crux of the matter is that if shareholders are slightly 'ferromagnetic', stock prices can then drop sharply, thus meeting the following well-known financial rule: "market needs \emph{money} for shares to go up, yet it needs nothing for them to drop". In magnetocaloric terms, we would say that "for the temperature of the magnetic material to rise, we need \emph{energy}, yet we need nothing to cool it".
\subsection{Limitations of the temperature signing algorithm}
\begin{figure}[ht]
\begin{minipage}[b]{0.45\linewidth}
\centering
\includegraphics[width=\textwidth]{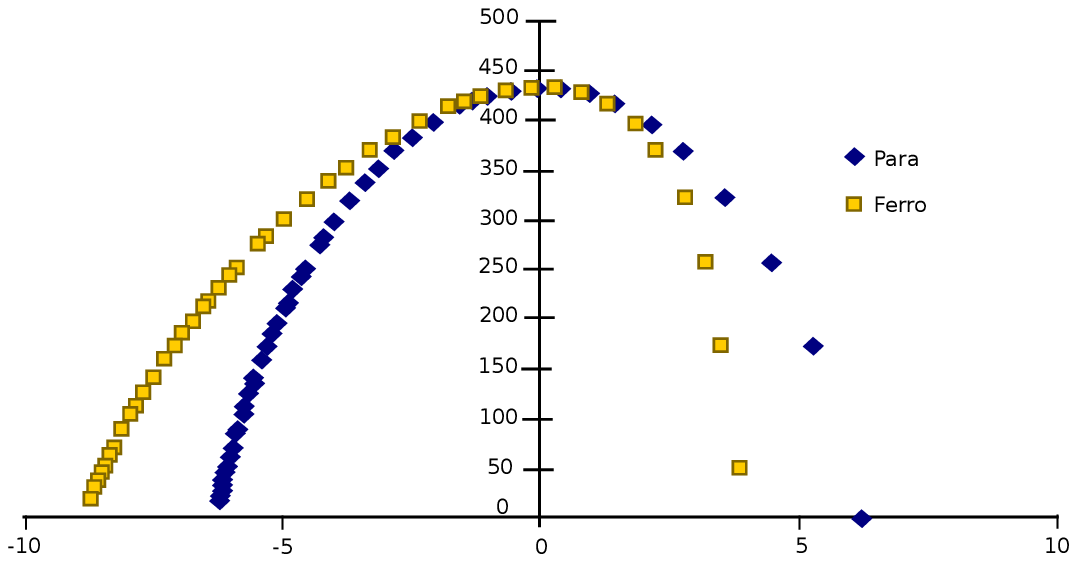}
\caption{Monte Carlo simulation for $J=0$(paramagnetism) and $J\approx 0$(slight ferromagnetism) of spin system $j=1/2$. Slightly ferromagnetic behaviour can be observed to move away from ideal paramagnetic behaviour.}
\label{ParaFerro}
\end{minipage}
\hspace{0.5cm}
\begin{minipage}[b]{0.45\linewidth}
\centering
\includegraphics[width=\textwidth]{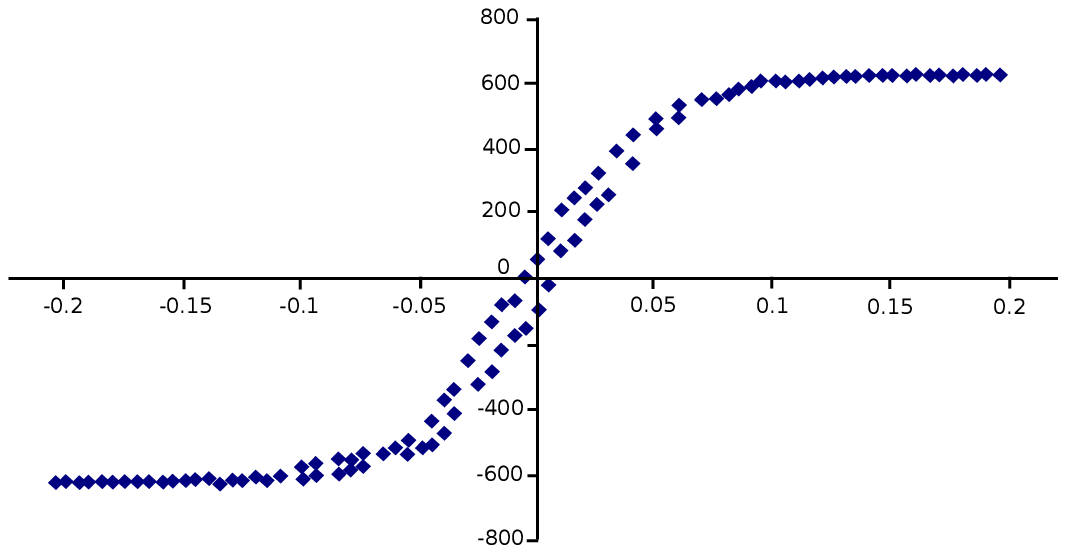}
\caption{Hysteresis loop Monte Carlo simulation for $J\approx 0$(slight ferromagnetism) of spin system $j=1/2$. A small part of the magnetic energy is lost.}
\label{Histeresis}
\end{minipage}
\end{figure}

Given the presence of random noise in the historic series of stock market prices, it would not be reasonable for a temperature signing algorithm to be based on one only method. From a probabilistic viewpoint, sign determination would be more reasonable through two different ways. If the conditions (15) and (16) in \cite{Subi13} based on equation (13) in \cite{Subi13} are taken as the first way, the second could be the rising trend in the absolute value of temperature due to the instability inherent to the states in which magnetic energy per spin $E$ is over zero, (17) in \cite{Subi13}.
Figure ~\ref{ParaFerro} represents a Monte Carlo simulation in which it can be observed how the entropic curve deforms due to slight ferromagnetism.
At the same time, the hysteresis loop is increased, thus partially losing magnetic energy (Fig.~\ref{Histeresis}).
For strong ferromagnetism levels, deformation is so remarkable that the criterion of the entropic curve slope (15) and (16) in \cite{Subi13} is no longer valid to determine the sign of the temperature (although the growing volatility criterion (17) in \cite{Subi13}, based on the Black-Scholes equation, remains valid).
\section{Scale invariance and temperature renormalization}
\begin{figure}[ht]
\begin{minipage}[b]{0.45\linewidth}
\centering
\includegraphics[width=\textwidth]{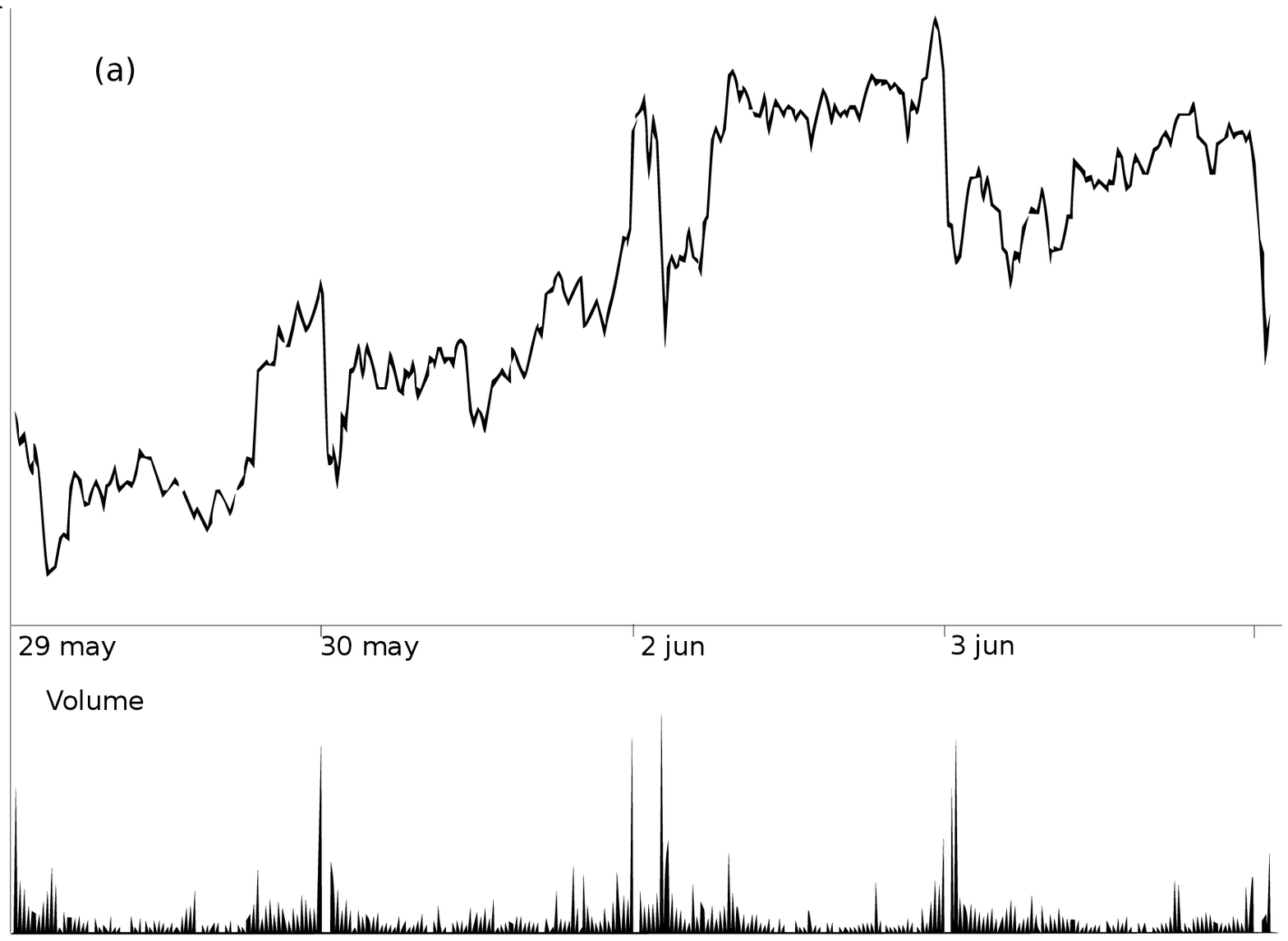}
\end{minipage}
\hspace{0.5cm}
\begin{minipage}[b]{0.45\linewidth}
\centering
\includegraphics[width=\textwidth]{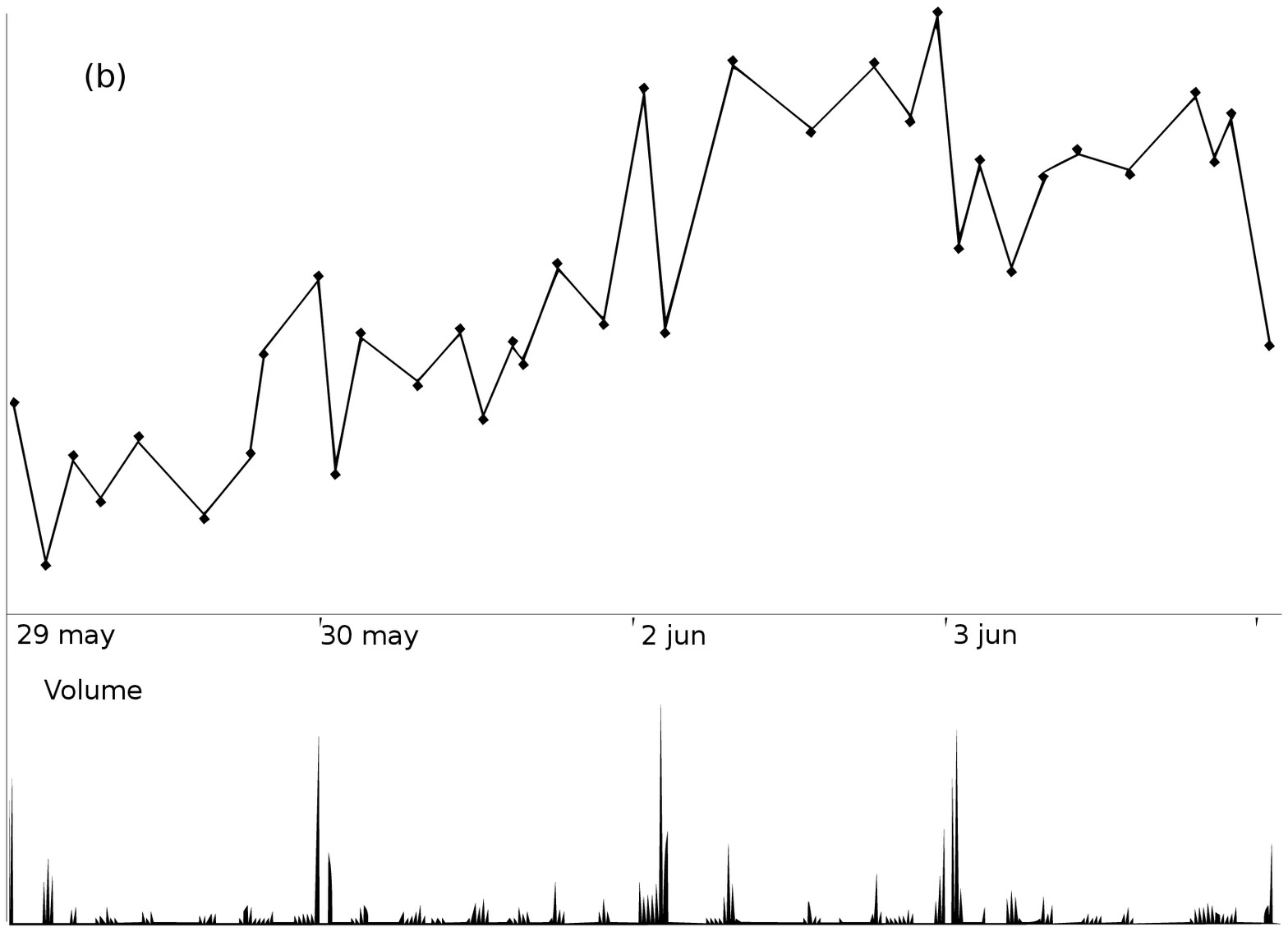}
\end{minipage}
\caption{(a)Graphic representation of the time series of stock intraday prices and (b) the same time series after renormalization.}
\label{Renormal}
\end{figure}

Numerous studies published in scientific journals support the thesis on the fractal nature of stock market prices. The description itself by R. N. Elliott of his own theory is a perfect example of fractality. The most striking aspect is that this description had already been made from an empirical viewpoint many years before (1939) Mandelbrot got his works published and defined the concept of fractality. If we gave an expert financial analyst a stock market graph in which units from X and Y axes have been removed, this analyst could not determine whether this graph is intradiary, diary, weekly or monthly. This scale invariance in the field of time is therefore unquestionable. The subsequent issue is how this affects stock market dynamics. The spin system model proposed in \cite{Subi13} may allow two perfectly differentiated phases with a phase transition at a hypothetical Curie temperature. This happens to fit in with experience: financial analysts distinguish two clearly different phases (uptrend and downtrend), and transition periods characterised by high volatility and lateral movement in stock prices.
On the other hand, scale invariance leads us to think of the possibility of applying some kind of renormalization. Leaving aside profound considerations on the topological dimension of the spin network, it is easy to understand that space interactions become variations through time, thus existing some kind of correspondence between network topology and Brownian movement of stock prices. In other words, the observed scale invariance (in the time domain) of stock prices would be the consequence of some kind of topological invariance in the spin network.
Let use suppose we dispose of the following data (illustrated in Fig.~\ref{Renormal}): time series of intradiary stock prices, and frequency of interactions (transactions).
This series would be associated with an \emph{intraday temperature} given by (5) in \cite{Subi13}. Relative frequency could be compared with probability density of interactions occurring at a specific time point. We would empirically ascertain that this probability is remarkably higher in the opening and closing auctions (beginning and end of the day), as shown in this figure. Stock-price renormalization (in the time domain) would then consist on substituting the original intradiary time series by another series in which the set of intradiary values are substituted by opening and closing prices values weighted by the aforementioned relative frequency. Graphically, this would be equivalent to reducing the original series substituting whole sets of terms by their average value.
If the reduced and weighted (renormalized) time series is applied the same temperature formula (5) in \cite{Subi13}, a second renormalized \emph{interday temperature} will be obtained.
\subsection{Yang-Lee Zeros}
\begin{figure}[ht]
\begin{minipage}[b]{0.45\linewidth}
\centering
\includegraphics[width=\textwidth]{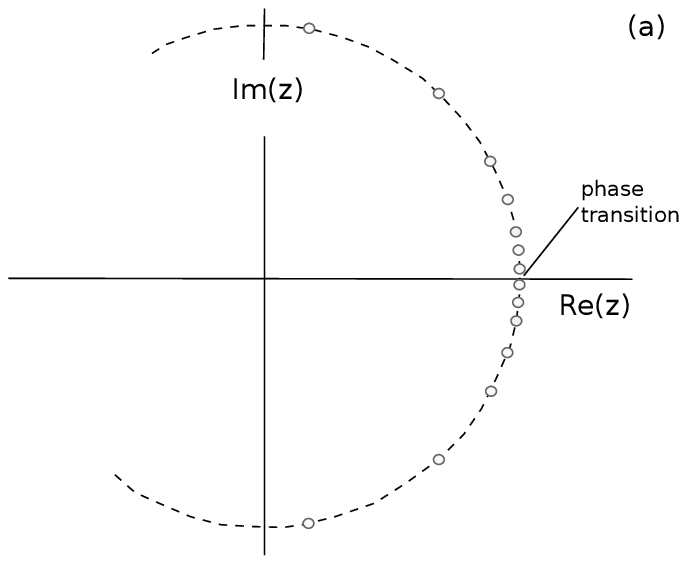}
\end{minipage}
\hspace{0.5cm}
\begin{minipage}[b]{0.45\linewidth}
\centering
\includegraphics[width=\textwidth]{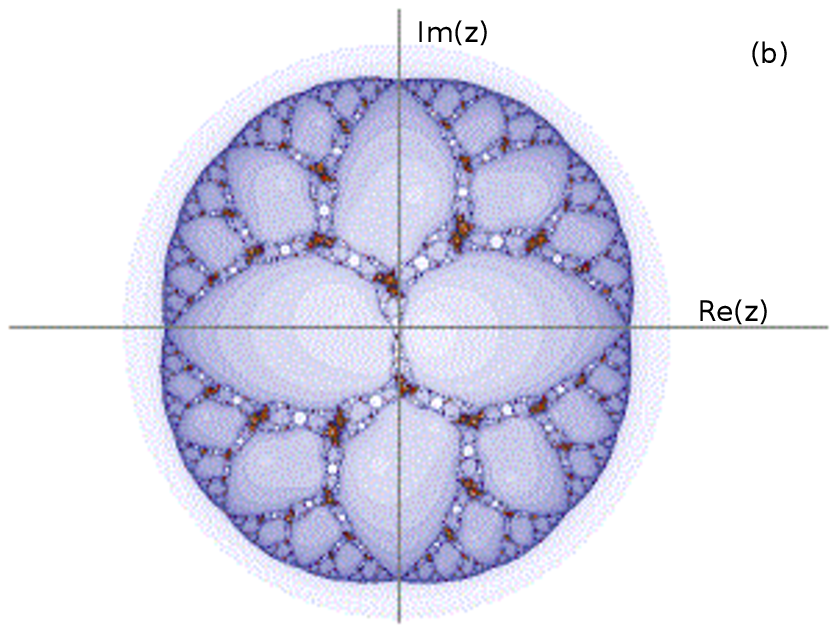}
\end{minipage}
\caption{(a) Accumulation of Yang-Lee zeros around the phase transition temperature; and (b) Map of zeros for some specific conditions.}
\label{YangLee}
\end{figure}

For spins organised in networks of simple topology, energy partition function ${Z_N}\left({T,H}\right)$ is a parametric function of temperature $T$ and external field $H$, $N$ being the number of particles. By means of renormalization, networks goes $N \to N'<N$, the number of degrees of freedom thus being reduced. Repeating the process $N<N'<N''<N'''...$ would lead us to $1$ or $2$ degrees of freedom, which is the trivial case, and we would have a series of subsequently renormalized temperatures $T<T'<T''<T'''...$, so that ${T^{(i+1)}}=R({T^{(i)}})$, where $R$ stands for the renormalization transformation. On the other hand, the partition function, transformed into polynomial form by means of a change of variable, has a set of zeros that correspond to points in the complex plane.
Years ago, \emph{Yang and Lee} proved phase transitions to be related to the distribution of zeros in the partition function \cite{Yang52-1,Yang52-2}. More precisely, as one gets closer to the thermodynamic limit, zeros begin to accumulate in a specific region of the complex plane and tend to 'mark' the real axis on the point corresponding to Curie temperature ${T_C}$ (Fig.~\ref{YangLee}~a), at which \emph{ferro-paramagnetic} phase transition occurs.
Furthermore, the map of zeros in the thermodynamic limit is equal to the \emph{Julia set} of the renormalization transformation. Thus, curious fractals such as the one shown in Fig.~\ref{YangLee}~b are obtained. However, the most interesting properties are that Curie temperature ${T_C}$ \emph{repels} the theoretical renormalization process, and that this temperature is \emph{invariant} relative to the renormalization transformation -i.e., ${T_C}=R({T_C})$. As an econophysic corollary of the latter property, it may be guessed that, for a financial market, if two or three successively-renormalized temperatures are calculated, downtrend-uptrend phase transition would then be found at those points with the same temperatures. In other words, phase transition could be identified when intraday temperatures get significantly closer to interday (renormalized) temperatures.
Figure 6 may confirm (although not statistically validated) this phase transition hypothesis, as it represents a well-known blue chip in Spanish IBEX 35. Arrows indicate hypothetical phase transition points in which temperatures values coincide and which also coincide with changes in price trends, thus marking trading opportunities.
\begin{figure}[tb]
\centering
 \includegraphics*[width=\textwidth]{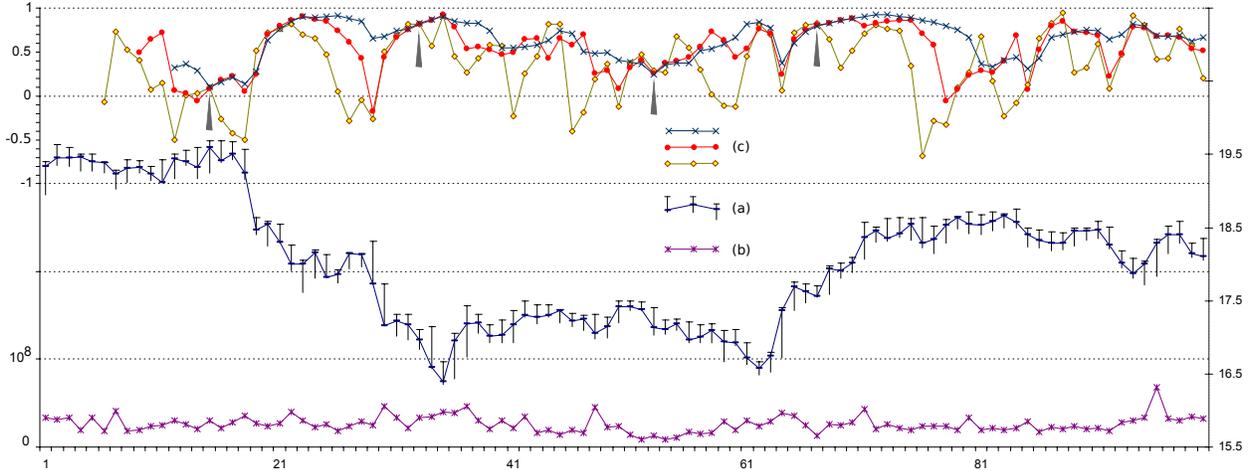}
  \caption{\label{Algotrader}(a) daily securities prices (End of Day) of a particular blue chip; (b) trading volumes; and (c) three successively-renormalized temperatures. Arrows indicate hypothetical phase transition points, which anticipate changes in trends.}
 \end{figure}
\appendix
\section{Trading strategies}
Nowadays important economic decisions are strongly influenced by powerful software tools. Thus, it can generally be said that a half-human, half-machine factor governs what some have called the modern cybernetic economy.
Getting into further detail, economic agents (institutional or retail investors, general public, etc.) can be observed to follow different trading strategies. Perhaps the most determining factor to choose a specific strategy is the frequency with which buying-selling cycles are to be completed, which in turns determines the mean lifetime and volatility of the assets in the corresponding investment portfolio. Anyway, the objective always remains the same: maximizing the portfolio's growth ratio.
The main current strategies are:
\begin{itemize}
  \item High Frequency Trading (HFT) basically consists on managing an investment portfolio whose assets' mean lifetime range from fractions of a second to some seconds. Since a human trader cannot take and perform decisions in such brief lapses of time, buying-selling orders are given by completely-automatic specialized algorithms, thus limiting the human factor to modulating or directing algorithm activity, as well as constant algorithm modification and enhancement. Thus, algorithms behave like an automaton that takes and perform decisions in fractions of a second. Low latency HFT -based exclusively on tiny price arbitrage advantages that render benefits of only a fraction of a currency cent- is a class of its own. A low latency algorithm opens and closes portfolio positions in a matter of microseconds and formal parallelism with a Brownian motor in physics is rather clear. In a more general context, automated or algorithmic trading is known as the use of software platforms that -preprogrammed with a set of rules or algorithms- send buying-selling orders to the market automatically. The set of preprogrammed routines is equivalent to a robot colloquially known as robotrader.
  \item Day trading: traders with technical analysis-based software tools complete buying-selling cycles on the same day -e.g., buying in the morning and selling in the evening. This strategy is exclusively based on technical analysis (Dow theory, Elliott waves, statistical indicators, etc.)
  \item Low Frequency Trading: this concept may include institutional or private investors that make use of fundamental analysis tools (macro- and micro-economic indicators, financial ratios, etc.) and whose investment portfolios are much less volatile than in the previous cases.
\end{itemize}
Algorithmic trading is widely used by large banks and institutional investors. The set of rules or algorithms can be based on any technique from technical or fundamental analysis to statistical physics. In $2006$, $30$\% of the negotiation volume in EU and US stock markets was estimated to be completed by algotraders or some other kind of automated trading. In $2009$, several studies suggested that HFT completed at least $60$\% of the total trading volume in the US. The prevalence of algotraders in some markets has even reached $80$\%. As a result, the massive use of these techniques is believed to have radically changed the market's microstructure and dynamics. Algorithmic trading in general was under debate until the US Securities and Exchange Commission informed that these techniques contributed to the wave of increased selling caused by $2010$ Flash Crash (a sudden crash in which Dow Jones plunged about $1000$ points in only some minutes' time, after which a large part of the loss was recovered).

\bibliography{SpinModel}

\end{document}